\documentclass[prb,preprint]{revtex4-1} 

\usepackage{amsmath}  
\usepackage{amsfonts} 
\usepackage{graphicx}
\usepackage{longtable}

\begin{document}

\preprint{APS/123-QED}

\title{A New Perspective on Impartial and Unbiased Apportionment}

\author{Ross Hyman}
\email{rhyman@uchicago.edu}
\affiliation{Research Computing Center, University of Chicago, Chicago IL 60637}
\author{Nicolaus Tideman}
\email{ntideman@vt.edu}
\affiliation{Economics Department, Virginia Tech, Blacksburg VA 24060}
\date{\today}

\begin{abstract}
How to fairly apportion congressional seats to states has been debated for centuries. We present an alternative perspective on apportionment, centered not on states but ``families'' of state, sets of states with ``divisor-method’’ quotas with the same integer part. We develop ``impartial" and ``unbiased" apportionment methods. Impartial methods apportion the same number of seats to families of states containing the same total population, whether a family consists of many small-population states or a few large-population states. Unbiased methods apportion seats so that if states are drawn repeatedly from the same distribution, the expected number of seats apportioned to each family equals the expected divisor-method quota for that family. 
\end{abstract}

\maketitle

\section{Introduction}
Every ten years, the U.S. House of Representatives is reapportioned according to the census for that decade. The Constitution specifies that ``Representatives shall be apportioned among the several States according to their respective numbers'' but prescribes no method to accomplish this. It is not a simple matter of multiplying the total number of seats in the House by each state's fraction of the population, since the resulting numbers of seats will generally not be integers. Many apportionment methods that assign each state an integer number of seats have been proposed, including methods by Thomas Jefferson, John Quincy Adams, Alexander Hamilton, and Daniel Webster \cite{B&Y,TJ,Malk}. Due to the political implications of the gain or loss of even one seat, disputes over apportionment methods have occurred many times, on the grounds that a proposed method is biased against states with small or large populations. The first such dispute resulted in George Washington's first veto \cite{B&Y}. The last such dispute, over the apportionment for the 1990 census, went to the Supreme Court \cite{Ernst}.

We present an alternative way to think about apportionment, in which the fundamental entities are not individual states but ``families'' of states, and we introduce ``impartial'' apportionment methods and ``unbiased'' apportionment methods. Impartial methods apportion seats among families of states according to the families' respective populations, regardless of whether a family contains a large number of small-population states or a small number of large-population states. For unbiased apportionment, we consider sets of states that are repeatedly drawn from the same distribution. In each family, the number of states and their populations, and therefore the family's total divisor-method quota, will, in general, differ across each sample set. Unbiased methods apportion seats so that if states are drawn repeatedly from the same distribution, the expected value of the total number of seats apportioned to each family equals the expected value of the total divisor-method quota for that family.

In sections \ref{history} and \ref{sliderule} we present a history of methods used or proposed for United States congressional apportionment, describe the Alabama paradox and the New States paradox, motivate apportionment methods using a ``slide rule,'' and show that divisor methods applied to states are not susceptible to the Alabama paradox or the New States paradox. 
In sections \ref{ImpartialApportionments} - \ref{SecNewState}
we define families of states and show that the Webster and Huntington-Hill methods applied to families are impartial and susceptible to the New States paradox. We find that Webster's method applied to families of states avoids the Alabama paradox while Huntington-Hill applied to families is susceptible to it. 
In sections \ref{IdealApportionments} - \ref{unbiasedlognorm} we consider state populations that are drawn from a specified probability distribution. We determine, for any probability distribution for populations, the apportionment method associated with that distribution that, when applied to states, is unbiased and avoids the Alabama and New States paradoxes. 

The link to a GitHub repository containing code for reproducing the tables in this paper and allowing for exploration of the apportionment methods described in this paper is in the references \cite{github}.

\section{A Brief History of Apportionment.}\label{history}
Article I, section 2 of the U.S. Constitution specifies:
\begin{quotation}
Representatives and direct Taxes shall be apportioned among the several States which may be included within this Union, according to their respective Numbers, . . . The actual Enumeration shall be made within three Years after the first Meeting of the Congress of the United States, and within every subsequent Term of ten Years, in such Manner as they shall by Law direct. The Number of Representatives shall not exceed one for every thirty Thousand, . . .
\end{quotation}
The Constitution does not specify a method of rounding the fractions that occur when apportioning integer numbers of representatives among states whose populations are not integer multiples of any feasible district size.

When Congress received the results of the 1790 census, it debated several methods of apportionment before settling on ``Hamilton's method,'' which involves deciding a size of the House, allocating seats including fractions to all states, rounding down to integers, and then apportioning the resulting extra seats to the states with the largest fractional remainders \cite{B&Y,Biles}. George Washington vetoed the bill that Congress passed, the only veto of his first term, on the ground that it gave Connecticut one representative for every 29,605 persons counted in the Census, in violation of the Constitutional requirement that the representation of each state not exceed one for each 30,000 \cite{B&Y,Biles}. Congress responded by adopting ``Jefferson's method,'' which involved dividing each state’s population by 33,000 and rounding down to an integer \cite{B&Y,Biles}. 

Jefferson's method was used for apportionments for all censuses through 1830, but Congress grew concerned over the method's bias in favor of large-population states. In the debate surrounding the apportionment for the 1830 census, other possibilities began to be discussed. John Quincy Adams proposed rounding up rather than down. James Dean, a mathematics professor from Vermont, proposed that the decision about whether a given state’s allotment should be rounded up or down should depend on which produced a ratio of persons per representative closer to the overall target, and Daniel Webster proposed ordinary rounding. For the 1840 census, Congress chose Webster’s method \cite{B&Y,Biles}.

For the apportionments from 1850 to 1900, Congress described the apportionment it chose as Hamilton's method, which had been their first choice in 1790, before Washington's veto. However, in 1850, 1870, 1880, and 1890, Congress deliberately chose a size for the House whereby Hamilton's method and Webster’s method yielded the same apportionment \cite{B&Y,Biles}.

For the apportionments of 1910 and 1930, Congress went back to Webster's method. (Congress failed to pass a new apportionment in response to the census of 1920.) For the apportionment in response to the 1930 census, Congress chose to keep the House size at 435, knowing that at that size, Webster’s method agreed with the Huntington-Hill method for the 1930 census. The Huntington-Hill method rounds down or up depending on whether a state’s ideally assigned number of seats, a real number that is generally not an integer, is less than or greater than $\sqrt{S(S+1)}$, where $S$ is the integer obtained by rounding the assigned number of seats down. Since 1940, Congress has chosen to use the Huntington-Hill apportionment method, keeping the size of the house at 435 seats \cite{B&Y,Biles}.

In 1880, when the Census Bureau calculated the allotments under Hamilton’s method for every House size from 275 to 350, Congress became aware that Hamilton’s method is subject to the ``Alabama paradox.'' This is the phenomenon that when the size of the House increases, it is possible for a state to lose a seat \cite{B&Y,Biles}. 

In 1907 Oklahoma became a state, and the 'New States paradox' was discovered when the Census Bureau calculated allotments using Hamilton's method. The New States paradox occurs when the size of the House increases through the addition of a new state along with additional seats appropriate for the size of the new state, and an existing state loses a seat \cite{B&Y,Biles}.

Michel Balinski and H. Peyton Young \cite{B&Y} identified apportionment methods that avoid both the Alabama paradox and the New States paradox. They also showed that if the populations of states are drawn from a uniform distribution, then a) the Huntington-Hill method \cite{H1,H2}, which Congress has used since 1930, is biased in favor of states with small populations and b) Webster's method does not favor states with small or large populations. In addition, they applied bias criteria to apportionments based on the Webster and Huntington-Hill methods and noted that the Webster apportionments showed less bias \cite{B&Y}.

For the 1990 census, the state of Massachusetts, which would have gained a seat with Webster's method, sued the U.S government on the grounds that the bias in favor of small-population states identified by Balinski and Young made the use of the Huntington-Hill method for congressional apportionment unconstitutional. The briefs in favor of the government's position were written by Laurence Ernst, a Census Bureau mathematician, who showed that for every criterion and measure of bias for which Webster apportionments were the least biased, there was a different criterion, for which the Huntington-Hill apportionments were least biased. The court decided in favor of the federal government, holding that, as there was no objective way to choose between the contending bias criteria, the choice between these contending methods was entirely subjective and within the purview of the Congress. Therefore, continued use of the Huntington-Hill method for congressional apportionment was constitutional \cite{Ernst}.

The problem of apportioning seats in a legislature to parties under a system of proportional representation is mathematically the same as the problem of apportioning seats in Congress among states. For a discussion of issues of apportionment in the context of proportional representation, see Pukelsheim \cite{Puk}.

\section{The Apportionment Slide Rule.}\label{sliderule}
\begin{figure}[!ht]
\centering
\includegraphics[width=\textwidth]{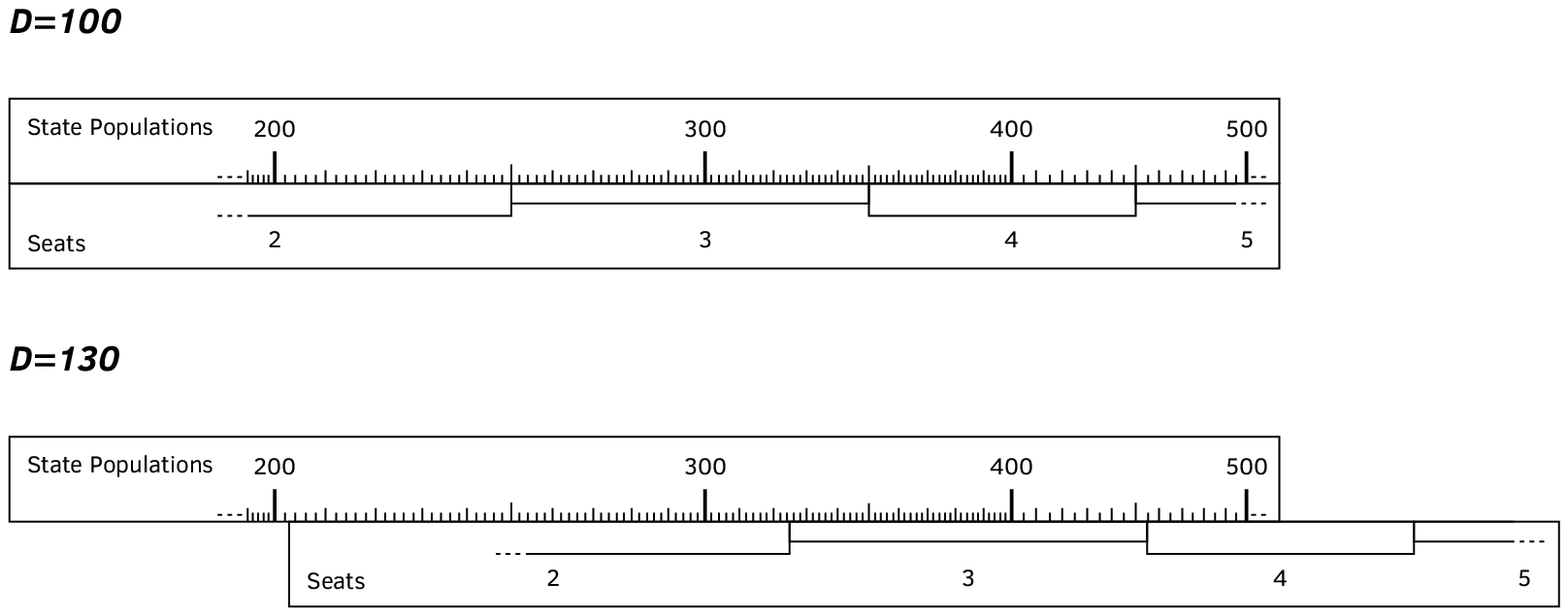}
\caption{A portion of the apportionment slide rule with half-integer rounding is shown at positions $D = 100$ and $D = 130$. Integer seat regions on the Seats ruler are shown with alternating thicknesses for clarity. For $D = 100$ a state with a population of 400 is apportioned 4 seats. At $D = 130$ it is apportioned 3 seats.}
\label{figslide}
\end{figure}
The integer number of seats apportioned to state $c$, $s_c$, would be perfectly proportional to the population of the state, $v_c$, if there were a parameter $D$ such that $s_c = v_c/D$ for every state. If the populations of states came only in multiples of the $D$, then a state would be assigned one seat for each $D$ persons, and there would be no apportionment problem. The apportionment problem arises because the ``quota'' for state $c$, $q_c = v_c/D$, is generally not an integer. We define the quota as the divisor-method quota in terms of $D$, which can be any positive real number. $D$ is the district size if seats are not rounded. It is not necessarily the average district size $v_T/s_T$ where $v_T$ is the total population and $s_T$ is the total number of rounded seats. The quota using $D$ and not the quota using $v_T/s_T$ must be used for the Alabama paradox to be avoided.

Apportionment can be understood in the following visual way. Figure \ref{figslide} shows two sliding rulers, one on top of the other, with logarithmic scaling on each. The top ruler is for state populations and the bottom ruler is for quotas and seats. Logarithmic scaling ensures that sliding the rulers corresponds to changing the common ratio, $D$, of population to unrounded seat quotas. For a given value of $D$, there is a relative positioning of the two rulers, such that the quota of seats allotted to a state with a given population is the number of seats on the Seats Ruler directly below the population of the state on the Populations Ruler. 

In general, quotas are not integers.  Most apportionment methods partition the Seats Ruler into integer seat segments. A rounding mark, $r(f)$, is placed on the Seats Ruler somewhere between integers $f$ and $f+1$, the precise position depending upon the apportionment method.  The segment of the Seats Ruler between consecutive rounding marks $r(f - 1)$ and $r(f)$ is the integer seat region for $f$ seats. When the State Populations Ruler is positioned over the Seats Ruler so that $v_c$ is over any part of the $f$ seat region, that is $r(f-1) \le q_c < r(f)$, state $c$ is assigned $f$ seats. 

A portion of the apportionment slide rule with half-integer rounding is shown in Figure \ref{figslide} at positions $D =100$ and $D = 130$. Integer seat regions on the Seats ruler are shown with alternating thicknesses for clarity. For $D=100$, states with 300, 400, and 500 people are apportioned 3, 4, and 5 seats respectively. At $D=130$, states with 300, 400, and 500 people are apportioned 2, 3, and 4 seats respectively.

We refer to the set of state quotas between $f$ and $f+1$, which are the quotas whose rounding is governed by $r(f)$, as the $f$ family of state quotas. A vision of an apportionment slide rule with a bell-shaped distribution of states on the Populations ruler inspired the ideas in this paper. As $D$ is increased, the population ruler is shifted to the left, with the logarithmic scaling preserving the shape of the distribution. Over each family on the Seats ruler, the density of states changes from sloping up to sloping down, as the left tail, followed by the middle, followed by the right tail of the distribution passes by. A rounding mark that cannot move in response to the changing shape of the density of states in its family tends to round up more of the smaller population states at the left tail of distribution and round down more of the large population states at the right tail of the distribution, producing an apportionment biased in favor of small population states. The rounding mark for each family must continuously adjust its position, in response to the changing shape of the distribution of states in its family, to ensure that the number of seats apportioned to the family remains as close as possible to the unrounded quota for that family.

In general, rounding marks can be dependent on $f$, $D$, and the state populations $\{v\}$, and can move if any of these change. These movements are the cause of the various paradoxes of apportionment. For example, if $D$ is decreased (increased) the State Populations Ruler moves to the right (left). If a rounding mark dependent on $D$ moves to the right (left) faster than the State Populations Ruler so that it outruns a state, then that state loses (gains) a seat, which is contrary to the motion of $D$, decreasing (increasing) the district size if seats are not rounded. This is the Alabama paradox. The total number of seats is adjusted by changing $D$, which corresponds to moving the State Populations Ruler relative to the Seats Ruler. For apportionment methods immune to the Alabama paradox, the total number of seats will not decrease (increase) when $D$ decreases (increases).

If a state is added or removed, while $D$ and the populations $\{v\}$ of the other states are held fixed, and a rounding mark dependent on state populations consequently moves through the population of an existing state, then that state's allotment will change. This is the New States paradox.

The Alabama paradox will never occur if $\partial\log r/\partial\log D \ge -1$ and the New States paradox will never occur if rounding for state $c$ can depend on $v_c$, $f$, and $D$ but not on the population of any other state. Both of these conditions are satisfied if $r(f)$ depends only on $f$ and not on $D$ or $\{v\}$. In other words, the rounding mark that determines the number of seats allotted to state $c$ depends only on the integer part of $q_c$ and not the quota of any other state. Such methods are called divisor methods \cite{B&Y}. A result of our paper is that there are methods that are immune to the Alabama paradox and the New States paradox yet are not divisor methods. 

\section{Families of States.}\label{ImpartialApportionments}
For both impartial and unbiased apportionment, we are concerned with the difference between the sum of the unrounded and rounded divisor-method quotas for a group of states. For impartial apportionment, we seek to minimize this difference while treating groups of states with the same total population equally. For unbiased apportionment we seek to make this difference precisely zero on average for states repeatedly drawn from the same distribution.  These two apportionment methods should agree as the number of states goes to infinity. But there is no hope that the difference between the sum of unrounded and rounded quotas approaches zero unless the groups include states whose quotas are rounded up and states whose quotas are rounded down.  

All the states whose quotas $q_c = v_c/D$ fall in the region $f\le q_c < f+1$, that is, all states with quotas whose integer part is $f$, we call the $f$ family of states. Each family contains one rounding mark. All states to the left of a rounding mark in a family are rounded down and those to the right are rounded up. So families are permissible groups for impartial and unbiased apportionment. In this paper we reexamine apportionment by treating families of states, rather than individual states, as the fundamental entities that are to be apportioned seats. 

Apportioning seats to a family as a whole uniquely determines the apportionment for the states in the family. To see this, consider the following method that apportions seats to families. We call the sum of the quotas of all states in the $f$ family $Q_f$. In general, $Q_f$ is not an integer. We round $Q_f$ to an integer following a prescribed apportionmnet method's rounding rule to determine the number of seats for the $f$ family, and we call that integer $S_f$. Once an integer $S_f$ is determined by the chosen method, the assignment of seats to the $N_f$ states in the $f$ family is uniquely determined by
\begin{eqnarray}
M_f &=& (f+1)N_f -S_f\nonumber\\
M_{f+1} &=& S_f - fN_f
\end{eqnarray}
where the $M_f$ smallest states in the family are assigned $f$ seats each and the $M_{f+1}$ largest states in the family are assigned $f+1$ seats each. Which states are in which group is uniquely determined by insisting that $s_c \ge s_d$ if $v_c > v_d$. 

Using this method, the rounding mark in the $f$ family is not explicitly needed or determined. It is any number from $f$ to $f+1$ that is smaller than the smallest state quota that is rounded up and larger than the largest state quota that is rounded down. 

Our analysis relies on two properties of families to determine impartial and unbiased apportionments that we were not able to use for other groups of states. One is that the boundaries of families are fixed and do not change when rounding marks change. The other is that each family can be treated independently of other families because each state in a family is rounded up or down depending on the position of its family's rounding mark and not that of any other family's rounding mark. 
\section{Impartial Apportionment.}\label{SecUnbiasedApportionment}
\begin{table}[!ht]
\centering
\caption{Differences between Webster for families and Webster for states.}
\begin{tabular}{l c  c c}
State&Quota&\multicolumn{2}{c}{Seats}\\
&& Webster for Families& Webster for States\\
\hline
North Dakota	&	1.024	&		1	&	1\\
South Dakota	&	1.166	&		1	&	1\\
Delaware &	1.302	&		1	&	1\\
Montana	&	1.426	&		1	&	1	\\
Rhode Island	&	1.443	&		2	&	1\\
Maine	&	1.791	&			2	&	2	\\
New Hampshire	&	1.812	&		2	&	2		\\
Hawaii	&	1.918 &	2	&	2\\
\hline
1 Family & 11.883	&	12 & 11\\
\hline
Louisiana	&	6.124	&		6	&	6		\\
Alabama	&	6.608	&			6	&	7		\\
South Carolina	&	6.733	&		7	&	7	\\
\hline
6 Family & 19.465	&	19	& 20\\
\hline
Minnesota	&	7.501	&		7	&	8		\\
Colorado	&	7.596	&		8	&	8		\\
Wisconsin	&	7.748	&		8	&	8		\\
\hline
7 Family & 22.846	&	23	& 24\\
\hline
Tennessee		&	9.087	&		9	&	9		\\
Massachusetts	&	9.240	&		9	&	9		\\
Arizona		&	9.405	&		10	&	9		\\
\hline
9 Family & 27.733	&	28	& 27\\
\hline
\end{tabular}
\label{ApportionmentTable}
\end{table}
Apportionment methods for families can be characterized by the rounding rule used to convert real $Q_f$ to integer $S_f$. In general, the rounding rule can depend on $Q_f$, $f$, $D$, and $\{v\}$. If we want the rounding rule to apportion the same numbers of seats to a family containing a large number of small-population states as it does to another family with the same total population, but containing a small number of large-population states, then the rounding rule can depend only on $Q$. Furthermore, we want the number of apportioned seats for the family to be as close as possible to $Q$ according to a measure of closeness such as Webster rounding which minimizes $|S-Q|$ or Huntington-Hill rounding which minimizes $|\log S - \log Q |$. We call an apportionment method that satisfies these properties, such as Webster's method applied to families and the Huntington-Hill method applied to families, an impartial method. Other divisor methods as well as Hamilton's method, can also be made impartial by applying them to families rather than states.

Rounding family quotas will produce the same apportionment as rounding state quotas when there is no more than one state in a family. However, when there is more than one state in a family, rounding $family$ quotas using any particular rounding rule may produce a different apportionment than rounding $state$ quotas using that same rule. This occurs because summing and then rounding will in general produce a different result than rounding and then summing.

Table \ref{ApportionmentTable} shows Webster for families (summing and then rounding) and Webster for states (rounding and then summing) apportionments using 2020 census data \cite{Census} for $D=v_T/435$. For each apportionment, the total number of apportioned seats is 435. The apportionment is shown for families 1, 6, 7, and 9 which are the families where the two apportionments differ. 

To make the total number of seats in each family as close as possible to its family quota, Webster for Families apportions an additional seat to Rhode Island and Arizona and one fewer seat to Alabama and Minnesota than Webster for States.  For each of the families involved, one can see that the difference between the Webster for families and Webster for states is due to a skewed distribution of states within those families. In the 1-Family, five states have quotas less than 1.5 while three states have quotas above 1.5. In the 6-Family, one state has a quota less than 6.5 and two states have quotas above 6.5. In the 7-Family, all three of its states have quotas above 7.5. In the 9-Family, all three of its states have quotas less than 9.5. 
\section{The Alabama Paradox and Impartial Apportionment.}\label{SecAlabama}
Divisor methods applied to families have many analogous properties to divisor methods applied to states. For example, the optimization functions discussed by Huntington \cite{H1,H2} and Balinsky and Young \cite{B&Y} when applied to families instead of states, are optimized at fixed $D$ by their respective traditional divisor methods applied to families instead of states. However, all divisor methods applied to states are immune to the Alabama paradox but not all divisor methods applied to families are immune. This is because unlike states, families can change their population size as $D$ is changed. This occurs when a state that is in one family for one value of $D$ is in a different family for another value of $D$. There is no counterpart to $D$ dependent population change for divisor methods applied to states, since the population of a state remains the same as $D$ is changed. 

When $D$ is decreased, and therefore the State Populations Ruler is slowly moved to the right, a state leaves the $f$ family and enters the $f+1$ family just as its state quota is the integer $f+1$ at the boundary between the $f$ and $f+1$ families. The family quota, $Q_f$, for the $f$ family decreases by $f+1$, an integer amount, and the family quota, $Q_{f+1}$, for the $f+1$ family increases by $f+1$. This can have effects on the other states in the two families if the rounded family quota, $S$, for a family changes by a different amount than the unrounded family quota, $Q$, when a state enters or leaves the family. This can happen if the rounding up or down of $Q$ depends on the integer part of $Q$, as is the case for Huntington-Hill rounding. It cannot happen if the rounding up or down of $Q$ does not depend on the integer part of $Q$, as is the case for Webster rounding.

Consider two states, one with quota 0.999 and the other with quota 1.43. Each is the only state in its family, the 0 and 1 families respectively. Using Huntington-Hill $\sqrt{S(S+1)}$ rounding on the family quota $Q$, the first state's 0.999 quota is rounded up to 1 and the second state's 1.43 quota is rounded up to 2. As $D$ is decreased so that every state's quota gets magnified by 1001/999, the first state's quota is now 1.001 and the second's quota is now 1.433. Now the two states are both in the 1 family with quota 2.433 which is rounded to 2. Therefore, the first state gets one seat as before, but now the second state also gets one seat. The second state lost a seat even though $D$ decreased. An apportionment method that can apportion fewer seats to a state when $D$ is decreased is susceptible to the Alabama paradox. To see this, add a third state with quota 999. Initially the total number of seats is 1002. After decreasing $D$ the third state's quota is 1001 and the total number of seats is 1003. The total number of seats has increased yet the second state has lost a seat. This is the Alabama paradox. 

If the rounding down or up of $Q$ depends on the integer part of $Q$, as with Huntington-Hill rounding, then the total number of seats can go down, up, or stay the same as $D$ decreases. In the above example, if the two states are the only states in the country, then the total number of seats went down from 3 to 2 as $D$ decreased. For an example in which the total number of seats remains the same, add a third state with quota initially 62.4375 which rounds down to 62 and the total number of seats is 65. After magnification by a factor of 1001/999, the third state's quota is 62.5625 which rounds up to 63 and the total number of seats is also 65. In other words, if Huntington-Hill rounding is applied to family quotas to allocate 65 seats among three states with relative populations of 0.999, 1.43, and 62.4375, the allocation can be achieved either with the original quotas, yielding an apportionment of 1, 2, and 62, or by quotas scaled by 1001/999, yielding an apportionment of 1, 1, and 63. This “multiple solution paradox” has not been mentioned previously in the literature because it cannot happen with apportionments by state. For an example where the total number of seats goes up when the seats apportioned to one of the states goes down, add a fourth state with quota 999. 

For conventional Webster rounding at 0.5, the first and second state are allotted one seat each, before and after $D$ is changed, and no Alabama paradox occurs. This is a general result. The Alabama paradox, including the multiple solution paradox, cannot occur for Webster's method applied to families. This is so because if $D$ is changed so that a state enters or leaves a family, the family quota changes by an integer amount exactly equal to the quota of the state that entered or left the family. For Webster's method, the direction of the rounding of the family quota does not depend on the integer part of the family quota so the direction of the rounding is unchanged when a state enters or leaves a family. The rounded family quota, like the unrounded quota, is changed by an integer amount exactly equal to the quota of the state that entered or left the family. Therefore, the seats apportioned to states in the family are unchanged and no Alabama paradox or multiple solution paradox occurs for Webster’s method applied to families.

If combining states into families with the same integer part of their state quotas is worth considering, why not also combine families together that have the same integer part of their family quota, and so on? One disadvantage of combining families is that such methods can violate the Alabama paradox, and therefore can have more than one apportionment for the same total number of seats. This happens because rounding then summing is different than summing then rounding. To see this, consider three state quotas: 0.99999, 1.7, and 2.6. They are each in a different family. And they are rounded to 1, 2, and 3. Decrease D so the 0.99999 state quota gets moved to 1.0 and is now in the 1 family. The quotas of the 1 family and the 2 family are now approximately 2.7 and 2.6 respectively. If we stop here, the family quotas round to 3 and 3, the states are assigned 1, 2, and 3 seats respectively, and no Alabama paradox has occurred. If we combine the 1 and 2 families into a family of the families with family quotas that round down to 2, then, this family-of-families has a quota of $2.6+2.7=5.3$, which rounds down to 5. The two families which compose the family-of-families are assigned 3 seats and 2 seats respectively, and the states are assigned 1, 2, and 2 seats. The Alabama paradox has occurred because a state has lost a seat when D has decreased. 

\section{The New States paradox and Impartial Apportionment.}\label{SecNewState}
Webster applied to families produces the least biased apportionment for a particular set of states, where the least biased apportionment is defined as the apportionment that minimizes $|S_f-Q_f|$ for every family. The method's rounding marks are determined impartially, dependent only on $Q$ and therefore $\{v\}$. On the other hand, rounding marks of apportionment methods that are immune to the New States paradox cannot depend on $\{v\}$. 

For example, a state with quota $2.6$ that is the only state in the 2 Family is apportioned 3 seats and a state with quota $5.3$ that is the only state in the 5 family is apportioned 5 seats. Add a state with quota $2.7$ without changing $D$. The new family quota of the 2 family is $5.3$ which rounds down to 5. Webster for families impartially apportions 5 seats to the 2 family and 5 seats to the 5 family. The $2.6$ state is apportioned $2$ seats and the $2.7$ state is apportioned $3$ seats. The $2.6$ state has lost a seat to maintain impartiality between one large population state with a quota of $5.3$ and two small population states with total quota $5.3$. If one instead apportioned in a manner immune to the New States paradox, a state with quota $5.3$ is rounded down to 5 and states with quotas $2.6$ and $2.7$ are rounded up to 3 seats each, which is biased in favor of the small population states.  An apportionment method cannot be both impartial and immune to the New States paradox. 

\section{Unbiased apportionment.}\label{IdealApportionments}
 If the states are independently drawn from a known or theoretical distribution of states, then it is possible to base the apportionment method on the distribution of states rather than a particular sample. Using a distribution from which states are independently drawn, our task is to find the apportionment method that is unbiased in the sense that if states are repeatedly and independently drawn from the same distribution, then seats are apportioned such that $\left<S_f\right>$, the expected value of the number of seats apportioned to the $f$ family, is equal to $\left<Q_f\right>$, the expected value of the total $f$ family quota. Such an apportionment method, for which $\left<S_f\right>=\left<Q_f\right>$ for every $f$, we call an unbiased apportionment. 

Balinski and Young \cite{B&Y} developed implications of state populations drawn from a uniform distribution. We generalize Balinski and Young's work by developing implications of a variety of population distributions. We do not adopt Balinski and Young's homogeneity assumption. The homogeneity assumption is that multiplying all state populations by a scale factor will not change an apportionment for the same total number of seats. To demonstrate that this apparently reasonable assumption is too restrictive, consider an apportionment method designed to be unbiased for states independently drawn from a bell-shaped distribution of population sizes. A set of states that are all drawn from the left end of the distribution should have rounding marks to the right of the midpoints of integer intervals to achieve unbiased apportionments, while a set of states with the same relative populations that are independently drawn from the right end of the distribution should have rounding marks to the left of the midpoints of integer intervals to achieve unbiased apportionments, violating homogeneity. It is because of this dropping of the homogeneity assumption that unbiased apportionment methods can be immune to both the New States paradox and Alabama paradox and yet, in general, not be divisor methods.

Define $p(v)$ as the density function for state population, $v$. We treat $p(v)$ as defined for all positive real numbers, even though population must in fact be an integer.
Given a $D$, the average quota for the $f$ family is 
\begin{equation}
    \left<Q_f\right> = \int_{fD}^{(f+1)D}p(v)\frac{v}{D}d v,
\end{equation}
and the expected value of the number of seats for the $f$ family is 
\begin{equation}
    \left<S_f\right> = f\int_{fD}^{rD}p(v)d v + (f+1)\int_{rD}^{(f+1)D}p(v)d v.
\end{equation}

An apportionment is unbiased within the $f$ family if 
\begin{eqnarray}
    \left<S_f\right> - \left<Q_f\right> &=& f\int_{fD}^{rD}p(v)d v + (f+1)\int_{rD}^{(f+1)D}p(v)d v\nonumber\\ 
    &-& \int_{fD}^{(f+1)D}p(v)\frac{v}{D}d v =0.
\end{eqnarray}
In terms of the cumulative distribution function, $I(v)$, defined up to a constant by
\begin{equation}
    \frac{d I(v)}{d v} = p(v),
\end{equation}
 we have   
\begin{eqnarray}\label{SF-QF}   
\left<S_F\right>-\left<Q_F\right> &=& (f+1)I((f+1)D) - fI(fD)
- I(rD)\nonumber\\ &-& \int_{fD}^{(f+1)D}\frac{d I(v)}{d v}\frac{v}{D}dv\nonumber\\
   &=&\int_{fD}^{(f+1)D}\frac{d}{dv}\left(I(v)\frac{v}{D}\right)dv - I(rD)\nonumber\\ &-& \int_{fD}^{(f+1)D}\frac{d I(v)}{d v}\frac{v}{D}dv\nonumber\\
   &=& \frac{1}{D}\int_{fD}^{(f+1)D}I(v)dv - I(rD)=0.
\end{eqnarray}
The condition for an unbiased apportionment is
\begin{equation}\label{RD}
I(rD)= \frac{1}{D}\int_{fD}^{(f+1)D} I(v) d v
\end{equation}
so that
\begin{equation}\label{RF}
r(f,D) = \frac{1}{D}I^{-1}\left(\frac{1}{D}\int_{fD}^{(f+1)D} I(v)dv\right).
\end{equation}
 The dependence of the rounding marks on $D$ means that unbiased methods are not, in general, homogeneous divisor methods. 

Since the rounding marks for unbiased apportionment are determined by an assumed state population distribution from which the states are drawn and not the actual state populations, it will, in general, differ from impartial apportionment, which is determined by the actual state populations and not by an assumed state population distribution. An apportionment from an unbiased method will not in general minimize $|S_F-Q_F|$ for each sample drawn from the distribution, and therefore will not be an impartial apportionment. Likewise, Webster for families applied to sets of states repeatedly drawn from the same distribution will not in general satisfy $\left<S_F\right>=\left<Q_F\right>$. However, impartial and unbiased apportionments will tend to agree if the number of states is large enough that any particular sample of states closely resembles the distribution it is drawn from. 

\section{Unbiased apportionments and paradoxes.}\label{u-o-a-p}
The New States paradox will not occur if the rounding marks, which depend on the distribution from which the states are drawn, do not change when states are added to or removed from the sample. However, if properties of the distribution, such as its mean and standard deviation, are determined from the sample, and they change when states are added to or removed from the sample, then changing the sample can change the rounding marks. Determining properties of a distribution from previous censuses while incorporating a model to update those parameters to the current census year is immune to the New States paradox, as long as the parameters are not further updated in response to the inclusion of the new state, but the unbiasedness of such an apportionment is only as accurate as the model.

The Alabama paradox will not occur if rounding marks cannot catch up with and overtake state populations when the State Populations Ruler is moved. As the State Populations Ruler moves to the right, $D$ decreases. Provided that $\frac{d\log r}{d\log D} \ge -1$, rounding marks do not move faster than the State Populations Ruler, and the Alabama paradox will not occur.

Taking the derivative of Eq.(\ref{RD}) with respect to $D$ we obtain
\begin{equation}
p(rD)D \frac{d rD}{dD} = \frac{1}{D}\int_{fD}^{(f+1)D}p(v)vdv.
\end{equation}
For this to be true
\begin{equation}
    \frac{d rD}{dD}\ge 0,
\end{equation}
which implies
\begin{equation}
    \frac{d \log r}{d \log D}\ge -1,
\end{equation}
and the Alabama paradox can never occur.

We believe that unbiased apportionment methods are the first example in the literature of apportionment methods that avoid the Alabama paradox and the New States paradox, yet are not divisor methods because they are not homogeneous.

Power law distributions and lognormal distributions are of particular interest as candidates for distributions from which states are drawn as there are numerous generative models for which state populations, as well as many other quantities, can be effectively treated as if they were drawn from such distributions \cite{Mitz}.

In the next section we show that many of the familiar divisor methods are unbiased apportionment methods for power law distributions. 

\section{Power law distributions.}\label{powerlaws}
\begin{table}[!ht]
\centering
\caption{Rounding Marks for Power Law Distributions}
\begin{tabular}{l c c c c c c c c c c c}
$f$& \multicolumn{11}{c}{$\beta$} \\
&$-\infty$&-4&-3 &-2&-1&0&+1&+2&+3&+4&$+\infty$\\
\hline	
0&0.00&	0.00	&0.00	&0.00&	0.00&	0.37&	0.50	&0.58	&0.63	&0.67&1.00\\
1&1.00&	1.36&	1.39&	1.41&	1.44&	1.47&	1.50&	1.53	&1.55	&1.58&2.00\\
2&2.00&	2.42&	2.43&	2.45&	2.47&	2.48&	2.50&	2.52&	2.53&	2.55&3.00\\
3&3.00&3.44	&3.45&	3.46&	3.48&	3.49	&3.50	&3.51	&3.52	&3.54&4.00\\
4&4.00&	4.45&	4.46	&4.47&	4.48	&4.49	&4.50	&4.51	&4.52&	4.53&5.00\\
\end{tabular}
\label{RoundingRules}
\end{table}
Consider the power law population distribution, $p_\beta\propto v^{\beta-1}$. The cumulative distribution function is $I(v) \propto v^\beta$. Solving Eq. (\ref{RF}) for $r_\beta(f)$ we have
\begin{equation}
r_\beta(f)= \frac{1}{D}\left(\frac{1}{D}\int_{fD}^{(f+1)D} v^{\beta}dv\right)^{1/\beta}= \left(\frac{(f+1)^{\beta+1} -f^{\beta+1}}{\beta +1}\right)^{1/\beta}.
\end{equation}
The rounding mark $r_\beta(f)$ does not depend on $D$, so power law apportionment methods are homogeneous divisor methods. They are the only unbiased apportionment methods with this property. This is so because power law distribution functions are the only distributions that have the same shape on any interval.

Examples of $r_\beta$ are shown in Table \ref{RoundingRules}. Some values are determined by taking the appropriate limits. Formulas for some of the values are shown below.
\begin{eqnarray}
&&r_{-\infty}(f) = f,\\
&&r_{-2}(f)  = \sqrt{f(f+1)},\\
&&r_{-1}(f) = \left(\log(f+1) - \log(f)\right)^{-1},\\
&&r_0(f) = \frac{(f+1)^{f+1}}{ef^f},\\
&&r_1(f) = f + \frac{1}{2},\\
&&r_2(f) = \sqrt{f(f+1) +\frac{1}{3}},\\
&&r_\infty(f) = f+1.
\end{eqnarray}
Several of these rounding rules have special names. $r_{-\infty}$ is the Adams apportionment rounding rule.  $r_{-2}$ is the Huntington-Hill rule.  $r_1$ is the Webster rule. $r_{\infty}$ is The Jefferson rule. There are some divisor methods, such as Dean's method, that are not derivable from any power law.

As Balinski and Young noted \cite{B&Y}, if the states are drawn from a uniform distribution, Webster's method is unbiased. 
However, if the states are drawn from a different distribution, then Webster's method will be biased. 

In the next section we consider unbiased apportionments for states drawn from lognormal distributions.

\section{Lognormal Distributions.}\label{unbiasedlognorm}
\begin{figure}[!ht]
\centering
\includegraphics[width=\textwidth]{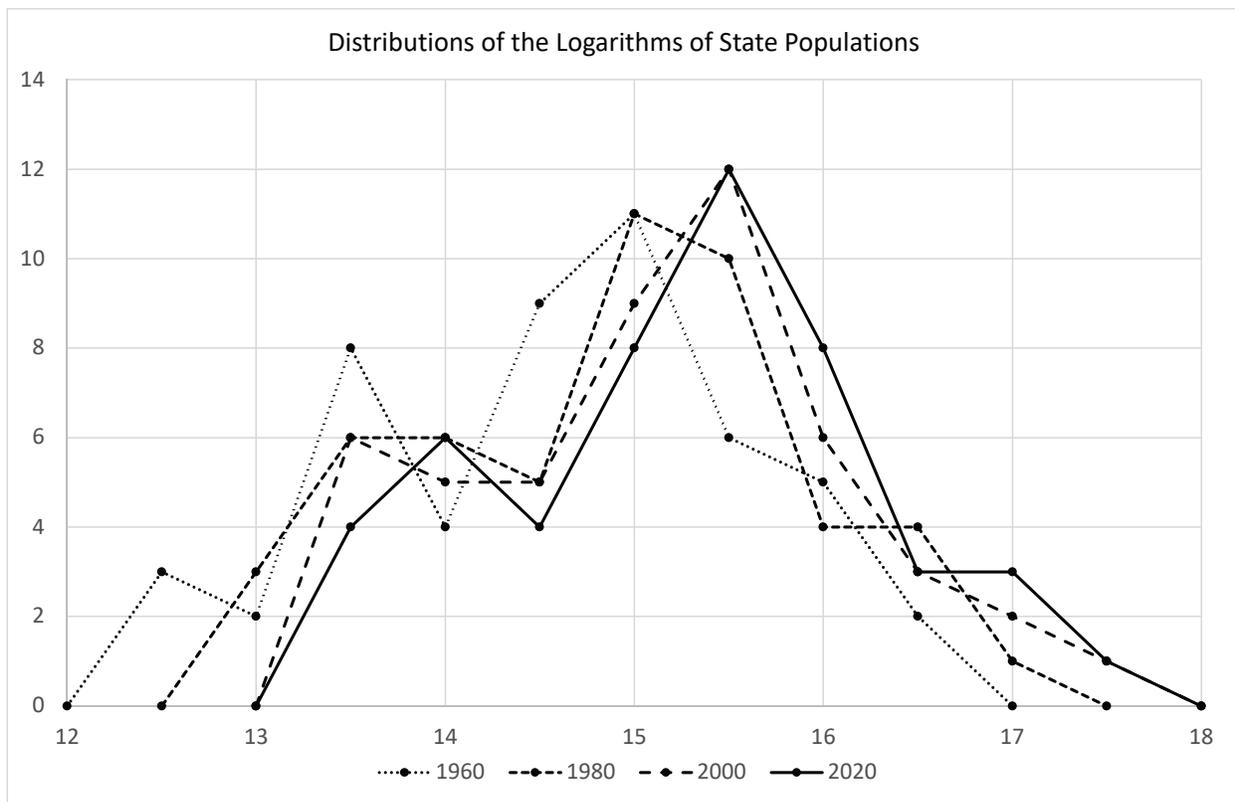}
\caption{Histograms of the distributions of the logarithms of state populations are shown for recent censuses at 20-year intervals.}
\label{fig1}
\end{figure}

Figure \ref{fig1} shows histograms of the distributions of the logarithms of state populations for recent censuses \cite{Census} at 20-year intervals.
 The distributions are roughly bell-shaped. There is no single power law distribution that can fit these distributions, since a positive exponent is required to fit small populations and a negative exponent is required to fit large populations.  The $\beta=0$ distribution is the power law distribution that fits census data best, in that it maximizes the likelihood of the observed distribution for integer $\beta$, but the distribution fails at the tails, which matter the most when considering size bias. 
\begin{table}[ht!]
\centering
\caption{Historical Statistics of Log($v$).}
\begin{tabular}{l c c c c}
Year&mean&standard deviation&skew&excess kurtorsis\\
\hline	
2020&15.218&1.024&-0.047&-0.514\\
2010&15.156&1.019&-0.054&-0.537\\
2000&15.062&1.020&-0.052&-0.572\\
1990&14.939&1.018&-0.015&-0.630\\
1980&14.850&1.021&-0.073&-0.719\\
1970&14.714&1.063&-0.091&-0.719\\
1960&14.583&1.071&-0.172&-0.606\\
\end{tabular}
\label{logvstats}
\end{table}
Table \ref{logvstats} shows moments for the logarithm of state populations for recent state population distributions \cite{Census}. The skew and excess kurtosis, both zero for normal distributions, are sufficiently small that the lognormal distribution is a reasonable approximation for these distributions. The mean of the logarithm of state populations increases almost perfectly linearly with time, with an $R^2$ value of $0.99$. The standard deviation of the logarithm of state populations appears to be decreasing towards a fixed point approximately equal to 1. 

We determine the unbiased rounding marks for the lognormal distribution from Eq. (\ref{RF}) by integrating and inverting its cumulative distribution function. Expressing the cumulative distribution function in terms of $q=v/D$, we have $I_{LN}(q,\log q_g , \sigma ) = I(qD,\log (q_gD),\sigma)$, where $q_g = v_g/D$, $\log q_g$ is the mean of $\log q$, and $\sigma$ is the standard deviation of $\log q$. The rounding mark for the $f$ family is determined from 
\begin{equation}
I_{LN}(r,\log q_g,\sigma)= \int_{f}^{f+1} I_{LN}(q,\log q_g,\sigma) d q,
\end{equation}   
which equals
\begin{eqnarray}
  &&I_{LN}\left(r,\log q_g,\sigma\right) = I_{LN}\left((f+1),\log q_g,\sigma\right)(f+1)\nonumber\\
  &&-I_{LN}\left(f,\log q_g, \sigma\right)f - I_{LN}\left((f+1),\sigma^2+ \log   q_g,\sigma\right)e^{\sigma^2/2}q_g \nonumber\\&&
  +  I_{LN}\left(f,\sigma^2 + \log q_g,\sigma\right)e^{\sigma^2/2}q_g.
\end{eqnarray}
 Provided that $v_g$ and $\sigma$ are determined a priori and not from the actual state populations, the apportionment method associated with these rounding marks is immune to the New States paradox and the Alabama paradox, yet it is not a divisor method. The rounding mark between $f$ and $f+1$ depends on $f$, as it would for a divisor method, but it also depends on $D$ through $q_g$, which indicates that it is not homogeneous. 

\begin{table}[!ht]
\centering
\caption{Rounding Marks for Lognormal Distributions.}
\begin{tabular}{l c c c c c}
$f$& \multicolumn{5}{c}{$q_g$} \\
&	1	&	2	&	5	&	10	&	20	\\
\hline	
0	&	0.491	&	0.539	&	0.591	&	0.623	&	0.650	\\
1	&	1.461	&	1.481	&	1.506	&	1.525	&	1.543	\\
2	&	2.468	&	2.480	&	2.495	&	2.507	&	2.518	\\
5	&	5.479	&	5.485	&	5.492	&	5.497	&	5.502	\\
10	&	10.487	&	10.489	&	10.493	&	10.496	&	10.499	\\
20	&	20.492	&	20.493	&	20.495	&	20.497	&	20.498	\\
\end{tabular}
\label{LNRoundingRules}
\end{table}

Table \ref{LNRoundingRules} shows rounding marks for lognormal distributions for various $q_g$ and $\sigma=1$. Looking at the $q_g=5$ column, one sees that the fractional part of a family's rounding mark starts above 0.5 and initially decreases with increasing family number, passing through 0.5 near the mode of the distribution, consistent with the idea that as one moves from family to family, from the left end of the distribution to the right end of the distribution, the fractional part of the rounding mark will be on the side of 0.5 where the distribution of states in a family is larger. As the family number continues to increase, the fractional part of the rounding mark continues to fall, eventually increasing again to approach 0.5, just as it does for power law distributions. Looking at the $f=5$ row, when $q_g=1$, the $5$ family is in the right tail of the distribution so its rounding mark is at the left side of the family. As $q_g$ increases, the rounding marks move to the right as the 5 family moves into the left tail of the distribution. 

The lognormal apportionment using the mean and standard deviation of the 2020 census populations \cite{Census} for which $q_g = 5.34$, does not agree with Webster's method applied to families. This is because the actual distribution of states has bumps at the tails that deviate from the lognormal distribution. The lognormal apportionment does agree with Webster applied to states for the 2020 census.

\section{Conclusion.}
We have presented impartial and unbiased methods for Congressional apportionment. Webster applied to families is an impartial apportionment method that apportions the same number of seats to families of states containing the same number of people regardless of whether a family is composed of a large number of small-population states or a small number of large-population states. Webster applied to families is immune to the Alabama paradox but not the New States paradox as no method can be both impartial and immune to the New States paradox. Unbiased methods apportion seats so that if states are drawn repeatedly from the same distribution, the expected value of the total number of seats apportioned to each family equals the expected value of the total divisor-method quota for that family.  Unbiased methods are immune to the Alabama paradox and the New States paradox yet not all of them are divisor methods. Unbiased methods for power law distributions are divisor methods. Unbiased methods for distributions that are not homogeneous, such as lognormal distributions, are not divisor methods. 

\begin{acknowledgments}
The authors thank the editor, editorial board, and reviewers for suggestions that improved the paper, and Carlos Reyes Zgarrick for his illustration of the apportionment slide rule.
\end{acknowledgments}

\end{document}